\documentclass[11pt]{article}

\usepackage{cite}
 \usepackage{subfigure}
\usepackage{multirow}
\usepackage{helvet}
\usepackage{amsmath}
\usepackage{amssymb}
\usepackage{setspace}
\usepackage{graphicx}
\usepackage{empheq}
\usepackage{soul}
\usepackage{tabularx}
\usepackage{array}
\usepackage{float}
\usepackage{braket}
\usepackage[utf8]{inputenc}
\usepackage{url}
\usepackage{hyperref}
\usepackage{slashed}
\usepackage[font=footnotesize,labelfont=bf]{caption}
\usepackage{color}

\def\be{\begin{equation}}
\def\ee{\end{equation}}

\usepackage[a4paper,top=3cm,bottom=3cm,left=2.5cm,right=2.5cm,bindingoffset=0mm]{geometry}

\begin{document}

\begin{center}

{\Large \bf Incoherent Particle Production in\\\vspace*{0.2cm}  Ultraperipheral  Heavy Ion Collisions}

\vspace*{1cm}
L. A. Harland-Lang \\                                          
\vspace*{0.5cm}                                                    
Department of Physics and Astronomy, University College London, London, WC1E 6BT, UK \\   

\begin{abstract}
\noindent In this paper, we present the first  complete treatment of incoherent photon--initiated production in ultraperipheral heavy ion collisions, focussing on the dilepton and diphoton final states. In the former case we compare to the ATLAS measurement of dielectron production and find that our predictions match the data very well. In the latter case we show that this  contribution is too small to explain the observed broad in diphoton acoplanarity background to the purely coherent light--by--light scattering signal. We in addition consider a new `quark emission' topology, which while naively might be expected to dominate the incoherent diphoton production channel, is in fact found to be kinematically suppressed. Finally, we revisit QCD--initiated diphoton production and issue of theoretical uncertainties in this case. We find that this again cannot explain the observed size of the higher acoplanarity background. The production mechanism leading to this background to light--by--light production, and the reason for its observed enhancement in comparison to the dilepton case, therefore remains unclear.

\end{abstract}

\end{center}

\section{Introduction}

The ultraperipheral collision (UPC) of heavy ions occurs when the impact parameter separation of the ions is much larger than the range of QCD, and hence no colour flow occurs between the colliding ions. In this environment, photon--initiated (PI) particle production is particularly enhanced, and this has enabled a rich set of experimental measurements at the LHC~\cite{ATLAS:2020epq,ATLAS:2020hii,ATLAS:2022srr,ATLAS:2022ryk,CMS:2022arf,CMS:2020skx,CMS:2024bnt,ATLAS:2025nac} and RHIC~\cite{STAR:2019wlg}.

The dominant PI production mechanism leading to UPC processes is due to purely coherent production, that is where both ions act coherently as a source of photons, and hence the cross section is enhanced by a factor of $Z^4$, where $Z$ is the ion electric charge. However, it is equally possible for either (or both) ions to interact incoherently, due to photon emission from the individual nucleons within ion, such that  the ion necessarily breaks up after the interaction. This contribution, in the dominant case where only one ion interacts incoherently, is expected to be suppressed by a factor of $\sim 1/Z$ and hence enter at the percent level with respect to the purely coherent case for typical LHC ion beams. While small, this may therefore not be negligible at the level of precision being aimed for at the LHC. Moreover the kinematics are in general rather different, with the incoherent mechanism dominantly leading to larger momentum transfers to the final--state system, as typically measured via the final--state acoplanarity, a proxy for the transverse momentum of the centrally produced system.

This incoherent production mechanism is particularly topical in light of the ATLAS measurement~\cite{ATLAS:2022srr} of dielectron production in UPCs, where the contribution from the higher acoplanarity tail is extracted via a template fitting technique. However, until now no model for such incoherent production has been available, in order to compare  these data with theoretical expectations. In this paper we therefore resolve this unsatisfactory state of affairs, by providing the first full account of incoherent PI production in UPCs, including a complete treatments of the initial--state photon flux, the final--state kinematics, and the survival factor probability of no additional ion--ion inelastic interactions. This will be made available in the \texttt{SuperChic} Monte Carlo (MC) generator. We will compare against the ATLAS data and find rather good agreement in general, validating the overall approach.

Beyond dilepton production, a further strong motivation for considering this issue is the case of light--by--light (LybL) scattering, as measured by both the ATLAS~\cite{ATLAS:2020hii} and CMS~\cite{CMS:2024bnt} collaborations. Here, the contribution to the fiducial region from larger diphoton acoplanarities (that is, larger diphoton transverse momenta), which is where the incoherent contribution is expected to dominate, is found to be rather large and of order the size of the coherent signal. In these analyses, this is assumed to be due to QCD--initiated production, namely via the $gg\to \gamma\gamma$ subprocess as part of an overall colour--singlet exchange between the beams. However, in a recent study~\cite{Klusek-Gawenda:2025ffh} it was suggested that incoherent PI production may instead be relevant here, with their estimates suggesting it may enter at the $\sim 20 \%$ level with respect to the coherent signal.

We will therefore apply the same approach developed in this paper to model incoherent dilepton production to the LbyL scattering case. We find that after accounting for all relevant effects omitted in the initial study of~\cite{Klusek-Gawenda:2025ffh}, due in particular to the kinematic cuts on the final--state photons, the incoherent PI contribution is expected to be small, at the sub--percent level. As we will discuss, this is significantly lower than the size of background observed by both ATLAS and CMS. 

With this in mind, we will also consider for the first time the possible contribution from photon emission directly from the quark line in the nucleon of the incoherently interacting ion. While naively enhanced by a factor of $\sim 1/\alpha^2$ in comparison to the `signal--like' incoherent  contribution, namely that which proceeds via loop--induced LbyL scattering, we find that accounting for all relevant  factors this is somewhat suppressed relative to it. This therefore cannot explain the sizeable background observed by ATLAS and CMS. Related to this, we briefly comment on the possibility of purely exclusive two--photon photoproduction off the nucleon. On general grounds this is expected to again be suppressed, being proportional to the valence generalized quark PDFs, although we raise the possibility of a related inelastic channel, a full analysis of which is left to future work.

Finally, we revisit the case of QCD--initiated production, for which the central predictions in~\cite{Harland-Lang:2018iur} were found to lie below the ATLAS and CMS extractions of the higher acoplanarity background by an order of magnitude or more. We will consider all possible sources of theoretical uncertainty but do not find any convincing possibility that these may explain the size of the background. The production mechanism leading to this background, and the reason for its enhancement in the LbyL case in comparison to the dilepton channel, therefore remains unclear, although in the concluding section we discuss some possible steps to shed light on this issue. 

The outline of this paper is as follows. In Section~\ref{sec:theory} we present the theoretical framework behind our approach, discussing in Section~\ref{sec:basic} the basic overview, in Section~\ref{sec:nuccs} the modelling of the nucleon--ion cross section, and in Section~\ref{sec:surv} the treatment of the ion--ion survival factor. In Section~\ref{sec:results} we present results. We start in Section~\ref{sec:pdf} with an approximate analysis within the photon PDF framework, and then in Section~\ref{sec:atlasdiel} present a comparison of the predictions within the full theoretical treatment to the ATLAS data on dielectron production. In Section~\ref{sec:lbylgam} we next consider the case of LybL scattering, starting in Section~\ref{sec:signallike} with the `signal--like' contribution, before considering the `quark emission' case in Section~\ref{sec:qe}, commenting on the possibility of purely exclusive two--photon photoproduction off the nucleon in Section~\ref{sec:photoproduction}, and finally revisiting the QCD--initiated case in Section~\ref{sec:QCD}. In Section~\ref{sec:conc} we summarise our results, with particular focus on the implications for the LbyL measurement.

\section{Theory}\label{sec:theory}

\subsection{Basic Framework}\label{sec:basic}

\begin{figure}
\begin{center}
\subfigure[]{\includegraphics[scale=0.6]{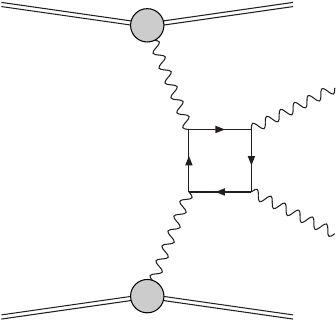}}\qquad
\subfigure[]{\includegraphics[scale=0.6]{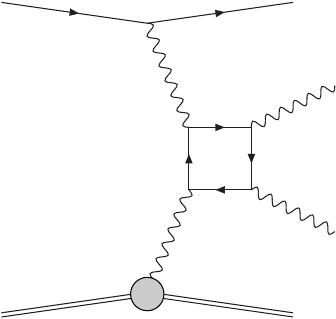}}
\caption{Representative Feynman diagrams for light--by--light scattering in AA ultraperipheral collisions in the case of (a)  coherent and (b) incoherent production. The lower double line corresponds to the ion beam that undergoes coherent photon emission, while the upper line corresponds to the incoherent interaction. The latter can be interpreted as either a nucleon or quark/anti--quark line depending on whether elastic or inelastic photon emission from the individual nucleons is considered.}
\label{fig:Feyn_incoh}
\end{center}
\end{figure}

For the case of incoherent PI production, the strongly dominant contribution is due to the mixed case where one ion continues to interact coherently; as we will see this contributes at the percent level to the cross section and hence production due to incoherent emission from both ions will enter at the sub per--mille level, and can be safely ignored. To first approximation, as in e.g.~\cite{Klusek-Gawenda:2025ffh} we have
\be\label{eq:aa_approx}
 \sigma_{AA}^{\rm incoh} = A   \,\sigma_{pA} + (Z-A) \, \sigma_{nA}\;, 
\ee
that is, the incoherent AA cross section is given in terms of the individual proton ($p$) and neutron ($n$) -- A cross sections, suitably summed over the nucleon content in the ion. A representative Feynman diagram for this $pA$/$nA$ cross section is shown in Fig.~\ref{fig:Feyn_incoh} (b), while the corresponding purely coherent case is shown in Fig.~\ref{fig:Feyn_incoh} (a) for comparison. For incoherent production, the photon emission from the nucleon can either be elastic or inelastic, exactly as in PI production in $pp$ collisions (see e.g.~\cite{Harland-Lang:2020veo}), and hence the  upper particle line in Fig.~\ref{fig:Feyn_incoh} (b) can either be interpreted as either a nucleon or quark/anti--quark line depending on whether elastic or inelastic emission is considered.

\subsection{Nucleon--ion cross section}\label{sec:nuccs}

As we will discuss further below, there are various reasons why \eqref{eq:aa_approx} requires modification, but nonetheless this provides the basic framework for the calculation. We are therefore interested in calculating the nucleon--ion cross section for the PI production of some system $X$. This can be done following the approach outlined in~\cite{Harland-Lang:2020veo,Bailey:2022wqy}, with the cross section given by
 \be\label{eq:sighhf}
 \sigma_{pA} = \frac{1}{2s}  \int  {\rm d}x_1 {\rm d}x_2\,{\rm d}^2 q_{1_\perp}{\rm d}^2 q_{2_\perp
} {\rm d \Gamma} \,\alpha(Q_1^2)\alpha(Q_2^2)\frac{1}{\tilde{\beta}} \frac{\rho_1^{\mu\mu'}\rho_2^{\nu\nu'} M^*_{\mu'\nu'}M_{\mu\nu}}{Q_1^2Q_2^2}\delta^{(4)}(q_1+q_2 - p_X)\;,
 \ee
 where the momentum fractions $x_{1,2}$ and $\tilde{\beta}$ are defined in~\cite{Harland-Lang:2019zur}, while $M^{\mu\nu}$ corresponds to the $\gamma\gamma \to X(k)$ production amplitude and ${\rm d}\Gamma = \prod_{j=1}^N {\rm d}^3 k_j / 2 E_j (2\pi)^3$ is the standard phase space volume. We have focussed on the $pA$ cross section for brevity, but the $nA$ cross section is given in the same way.  Here, $\rho$ is the density matrix of the virtual photon, which takes the general form:
 \be\label{eq:rho}
 \rho_i^{\alpha\beta}=2\int \frac{{\rm d}M_i^2}{Q_i^2}  \bigg[-\left(g^{\alpha\beta}+\frac{q_i^\alpha q_i^\beta}{Q_i^2}\right) F_1(x_{B,i},Q_i^2)+ \frac{(2p_i^\alpha-\frac{q_i^\alpha}{x_{B,i}})(2p_i^\beta-\frac{q_i^\beta}{x_{B,i}})}{Q_i^2}\frac{ x_{B,i} }{2}F_2(x_{B,i},Q_i^2)\bigg]\;,
 \ee
where $x_{B,i} = Q^2_i/(Q_i^2 + M_{i}^2 - m_p^2)$ for a hadronic system of mass $M_i$ and we note that the definition of the photon momentum $q_i$ as outgoing from the hadronic vertex is opposite to the usual DIS convention. Here, the integral over $M_i^2$ is understood as being performed simultaneously with the phase space integral over the photon momenta,  i.e. is not fully factorized from it. 

The inputs for the elastic and inelastic structure functions are described in Appendix~\ref{sec:sfinput}. However, here we discuss a modification that must be applied to the case of elastic proton emission (as discussed in the appendix we can safely drop the elastic neutron emission case) to account for the fact that the protons are bound inside the ions. It in particular should be the case that in order for incoherent emission from the ion to occur, the proton cannot scatter into an already occupied state in the ion, due to the Pauli exclusion principle. This `Pauli blocking' effect is typically modelled by requiring that the momentum transfer to the proton should be roughly larger than the typical Fermi momentum, $p_F \sim 250 $ MeV, of the nucleons in the ion (see~\cite{Bodek:2021trq} for a recent summary). More precisely, in the Fermi gas model this is accounted for by modifying the elastic proton form factors such that~\cite{Tsai:1973py}
\be\label{eq:pauli}
F_i^{\rm el, p/A}(x_{B,i},Q_i^2) =C_{\rm Pauli}(Q^2_i) \,F_i^{\rm el, p}(x_{B,i},Q_i^2)\;,
\ee
where $F_i^{\rm el, p}$ are the usual free proton elastic form factors and the modification factor $C(Q_i^2)$ is given by
\be\label{eq:pauli1}
C_{\rm Pauli}(Q^2)=\frac{3 Q_{\rm rec}}{4 p_F} \left[1-\frac{1}{12} \left( \frac{Q_{\rm rec}}{p_F}\right)^2\right]\;,
\ee
if $Q_{\rm rec} > 2 p_F$ and $C(Q_i^2)=1$ otherwise. Here 
\be
Q_{\rm rec}^2 =\left(\frac{Q^2}{2 m_p}\right)^2+Q^2\;.
\ee
This factor therefore reduces the corresponding elastic cross section by suppressing the region of lower photon $Q^2$. As the proton form factor is rather peaked towards low values of $Q^2$ this reduces the corresponding cross section by a factor of $\sim 5$, as we will discuss in the following sections. We note that in principle we could account directly for resonant excitation of the ions through photon emission, by suitably accounting for the inelastic ion form factors. However, as shown in~\cite{Hencken:1995me}, using a simple model for the dominant giant dipole resonant excitation of the ion, this contribution is expected to be at the per mille level with respect to the purely coherent cross section, and so we neglect it in what follows.

We note that, while \eqref{eq:sighhf}  is the general expression that we use, it is useful to make connect with simpler formulations when calculating the impact of the survival factor as well for the case where only the on--shell   $\gamma\gamma \to X$ amplitudes are available. To do this, taking the high--energy limit, and using the gauge invariance of the PI amplitude to drop the terms $\sim q_i$ in \eqref{eq:rho} we have
 \be\label{eq:rhosub}
  \rho_i^{\alpha\beta} =  2\int \frac{{\rm d}M_i^2}{Q_i^2}  \bigg[-g^{\alpha\beta} F_1(x_{B,i},Q_i^2)+ 2\frac{q_{i_\perp}^\alpha q_{i_\perp}^\beta}{Q_i^2}\frac{ x_{B,i} }{x_i^2}F_2(x_{B,i},Q_i^2)\bigg]\;.
 \ee
For elastic photon emission, the $F_2$ term dominates and we can write
\be\label{eq:csn}
\sigma =\frac{1}{2s}\int {\rm d} x_1 {\rm d}x_2 {\rm d}^2 q_{1\perp}{\rm d}^2 q_{2\perp} {\rm d}\Gamma\frac{1}{\tilde{\beta}} |T(q_{1\perp},q_{2\perp}) |^2\delta^{4}(q_1+q_2-k)\;,
\ee
where $T$ is the process amplitude, given by
\be\label{eq:tq1q2}
T(q_{1\perp},q_{2\perp}) = \mathcal{N}_1^p(x_1,Q_1^2) \mathcal{N}_2^A(x_2,Q_2^2) \,q_{1\perp}^\mu q_{2\perp}^\nu \mathcal{M}_{\mu\nu}\;,
\ee
with the normalization factors  
\be
\mathcal{N}_1^p(x_i,Q_i^2) = \frac{2\alpha(Q_i^2)^{1/2}}{ x_i}\frac{F_E(Q_i^2)^{1/2}}{Q_i^2}C_{\rm Pauli}^{1/2}(Q^2_i)  \;,\quad \mathcal{N}_2^A(x_i,Q_i^2) =\frac{2\alpha(Q_i^2)^{1/2}}{ x_i}\frac{F_p(Q_i^2)G_E(Q_i^2)}{Q_i^2}\;,
\ee
for the case that the incoherent (proton) beam is labelled $i=1$ and the ion is labelled $i=2$. The various form factors in the above expression are defined in Appendix~\ref{sec:sfinput}. In the full calculation we of course include both configurations, and note that when applying any cuts on the final--state system care must be taken to ensure that the appropriate boost is performed from the $pA$ to the $AA$ frame.

Finally, we note that, in the equivalent photon limit it was shown in~\cite{Harland-Lang:2010ajr} that \eqref{eq:tq1q2} can be recast via
\begin{align}
q_{1\perp}^\mu q_{2\perp}^\nu \mathcal{M}_{\mu\nu}=\begin{cases} &-\frac{1}{2} ({\bf q}_{1_\perp}\cdot {\bf q}_{2_\perp})(\mathcal{M}_{++}+\mathcal{M}_{--})\;\;(J^P_z=0^+)\\ 
&-\frac{i}{2} |({\bf q}_{1_\perp}\times {\bf q}_{2_\perp})|(\mathcal{M}_{++}-\mathcal{M}_{--})\;\;(J^P_z=0^-)\\ 
&+\frac{1}{2}((q_{1_\perp}^x q_{2_\perp}^x-q_{1_\perp}^y q_{2_\perp}^y)+i(q_{1_\perp}^x q_{2_\perp}^y+q_{1_\perp}^y q_{2_\perp}^x))\mathcal{M}_{-+}\;\;(J^P_z=+2^+)\\ 
&+\frac{1}{2}((q_{1_\perp}^x q_{2_\perp}^x-q_{1_\perp}^y q_{2_\perp}^y)-i(q_{1_\perp}^x q_{2_\perp}^y+q_{1_\perp}^y q_{2_\perp}^x))\mathcal{M}_{+-}\;\;(J^P_z=-2^+)
\end{cases}\label{eq:Agen}
\end{align}
where $\mathcal{M}_{\pm \pm}$ corresponds to the $\gamma(\pm) \gamma(\pm) \to X$ helicity amplitudes and $J^P_z$ the spin--parity of the corresponding $\gamma\gamma$ configuration. We will make use of this expression when calculating cross sections for processes where only the on--shell $\gamma\gamma \to X$ amplitudes are readily available, as is the case for LbyL scattering.

\subsection{Survival Factor}\label{sec:surv}

When calculating the cross section for incoherent UPC production we must account for the `no--overlap' probability that the colliding ions do not interact inelastically, as we do in the purely coherent case. This is as usual most clearly discussed in impact parameter space.  We therefore start with the cross section, written in impact parameter space, for the PI production of a system $X$ in $pA$ collisions (as before focussing on the proton case for brevity, but with it understood that the neutron case follows in a similar fashion). It can be shown that this is given by
\be\label{eq:pA}
 \sigma_{pA \to p A X} =  \int {\rm d}^2 b_{1\perp}{\rm d}^2 b_{2 \perp} \int \frac{ {\rm d} x_1 }{x_1} \frac{ {\rm d} x_2 }{x_2} |\tilde{\mathbf N}_p(x_1,b_{1\perp})|^2  |\tilde{\mathbf N}_A(x_2,b_{2\perp})|^2\,\sigma_{\gamma\gamma \to X}\;.
 \ee
Here
\be\label{eq:bflux}
 \tilde{\mathbf N}(x,b_{\perp})\equiv\frac{1}{(2\pi)} \int {\rm d}^2 q_{\perp} {\mathbf N}(x,q_{\perp}) e^{-i {\mathbf q}_{\perp}\cdot  {\mathbf b}_{\perp}}\;,
 \ee
 and
\be
{\mathbf N}_{p,A}(x_i,q_{i_\perp}) = \frac{x_i \mathcal{N}_1^{p,A}(x_i,Q_i^2)}{2\pi} \,{\mathbf q}_{i_\perp}\;.
\ee
To calculate the contribution from inelastic photon emission from a single ion, due to incoherent emission from the protons in the ion we then simply write 
\be\label{eq:AA}
 \sigma_{AA \to p A X}^{\rm inel, el} = \int {\rm d}^2 b_{1\perp}{\rm d}^2 b_{2 \perp}  \int {\rm d}^2 \tilde{b}_\perp T_{A,p}(\tilde{b}_\perp) \Gamma_{AA}(s,b_\perp') \frac{{\rm d}^2  \sigma_{pA \to p A X}}{{\rm d}^2 b_{1\perp} {\rm d}^2 b_{2\perp}}\;,
\ee
where $ {\mathbf b'}_\perp = {\mathbf b}_\perp-\tilde{{\mathbf b}}_\perp$, with ${\mathbf b}_\perp={\mathbf b}_{1\perp}+{\mathbf b}_{2\perp}$.
Here $T_A$ is the transverse proton density, given by
\be
T_{A,p}(b_\perp) = \int {\rm d} z \rho_p(r)\;,
\ee
where the proton density $\rho_p(r)$ is given in terms of the standard Woods--Saxon parameterisation. This is therefore normalized such that
\be\label{eq:TAnorm}
\int {\rm d}^2 b_{\perp}T_{A,p}(b_\perp)  = Z\;,
\ee
where $Z$ is the ion atomic number. In~\eqref{eq:AA},  $\Gamma_{A_1A_2}$ represents the usual probability that no inelastic scattering occurs at impact parameter $b_\perp'$. It is typically written in terms of the ion--ion opacity $\Omega_{A_1A_2}$ via
\be\label{eq:opac}
\Gamma_{A_1A_2}(s,b_\perp')  \equiv \exp(-\Omega_{A_1 A_2}(s,b_\perp'))\;.
\ee
This is given in terms of the opacity due to nucleon--nucleon interactions, $\Omega_{nn}$, which is in turn given by a convolution of the nucleon--nucleon scattering amplitude $A_{nn}$ and the transverse nucleon densities $T_n$, see~\cite{Harland-Lang:2018iur} for a more detailed discussion.

\begin{figure}
\begin{center}
\includegraphics[scale=0.7]{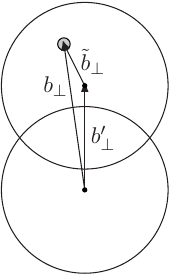}
\caption{Diagram indicating the definitions of the impact parameter vectors defined in~\eqref{eq:AA}.}
\label{fig:ionion}
\end{center}
\end{figure}

The physical origin of the additional integrals in~\eqref{eq:AA} is demonstrated in Fig.~\ref{fig:ionion}. In particular, the impact parameter between the proton that emits the photon and the centre of the other ion (which interacts coherently) is given by $ {\mathbf b}_\perp$, while the impact parameter between the centre of two ions is given by ${\mathbf b'}_\perp$, which is itself given in terms of $ {\mathbf b}_\perp$ and the transverse position $\tilde{{\mathbf b}}_\perp$ of the proton in the inelastically interacting ion. 

If we simply ignore the survival factor then the integral over $T_A$ simply drops out of \eqref{eq:AA} and we have
\be
\sigma_{AA \to p A X}^{\rm inel, el} = Z \,  \sigma_{pA \to p A X} \;,
\ee
using the normalization \eqref{eq:TAnorm}. That is, the total inelastic cross section simply becomes a sum over the individual $Z$ proton--ion interactions; for the neutron contribution we simply replace $T_{A,p}$  with the transverse neutron density, $T_{A,n}$, for which the normalization matches the $\sim (A-Z)$ prefactor as we would expect. This is the approach taken in~\cite{Klusek-Gawenda:2025ffh}. However we can see that this will certainly be an approximation that will overestimate the cross section. In particular, for certain impact parameter configurations it will not be the case that the entirety of the nucleons in the ion can take part in the interaction, due to the requirement that the ions themselves do not overlap, or more precisely interact inelastically. 

To account for this effect, we simply define
\be\label{eq:gaminelel}
\Gamma^{\rm inel,el}_{A_1A_2}(s,b_\perp)=   \int {\rm d}^2 \tilde{b}_\perp T_A(\tilde{b}_\perp) \Gamma_{AA}(s,b_\perp') \;,
\ee
in which case the cross section \eqref{eq:AA} becomes
\be
 \sigma_{AA \to p A X}^{\rm inel, el} = \int {\rm d}^2 b_{1\perp}{\rm d}^2 b_{2 \perp}  \Gamma^{\rm inel,el}_{AA}(s,b_\perp) \frac{{\rm d}^2  \sigma_{pA \to p A X}}{{\rm d}^2 b_{1\perp} {\rm d}^2 b_{2\perp}}\;,
\ee
i.e. it has the form as a standard proton--ion UPC cross section, but with a modified survival factor, and form factor on the inelastic side. This can therefore be readily translated to momentum space, and evaluated following the approach most recently outlined in~\cite{Harland-Lang:2023ohq,Harland-Lang:2025bkk}.
In particular, defining
\be\label{eq:pmc1}
\mathcal{P}^{\rm inel,el}(s,k_\perp)\equiv \frac{1}{(2\pi)^2}\int {\rm d}^2 b_{\perp}\,e^{i {\mathbf k}_{\perp}\cdot {\mathbf b}_{\perp}}\Gamma^{\rm inel,el}_{A_1A_2}(s,b_\perp)^{1/2}\;,
\ee
in terms of \eqref{eq:gaminelel}, then to account for the survival factor we simply replace \eqref{eq:tq1q2} with
\be
T_{{\rm S^2}}(q_{1\perp},q_{2\perp}) = \int  {\rm d}^2 k_{\perp}\, T(q_{1\perp}',q_{2\perp}')\,\mathcal{P}^{\rm inel,el}(s,k_\perp)\;,
\ee
where $q_{1\perp}' =q_{1_\perp} - k_\perp$ and $q_{2\perp}' = q_{2\perp} + k_\perp$. For the $F_1$ contribution, we consistently include the $F_1$ term in \eqref{eq:rhosub}, but add this incoherently with the dominant electric dipole contribution described above, see~\cite{Harland-Lang:2015cta} for further details. For inelastic photon emission from a nucleon we  apply the approach of the elastic case, but with the elastic form factors suitably replaced by the inelastic SFs. On physical grounds, we can expect a relatively smooth transition between the elastic and inelastic cases in the low $Q_i^2$, and by doing so we account for the physically relevant process and $q_{i_\perp}$ dependence of the survival factor. In principle for higher $Q^2$ values a more complete procedure should be pursued, as outlined in~\cite{Harland-Lang:2016apc}, however in reality this makes a small difference to the cross section predictions of interest, that is within other uncertainties, and hence we do not do this here.

 \section{Results}\label{sec:results}
 
 \subsection{First look: photon PDFs and survival factors}\label{sec:pdf}
 
As a first look, we can simplify the full expression~\eqref{eq:sighhf} in the high--energy and on--shell approximation, and neglecting the survival factor, to give
\be
 \sigma_{pA} \approx \int  {\rm d}x_1 {\rm d}x_2\, f_{\gamma/p}(x_1,\mu^2)\, f_{\gamma/A}(x_2,\mu^2)\hat{\sigma}(\gamma\gamma\to X)\;,
\ee
for the case that beam 1 interacts incoherently, and similarly for the case that the incoherent and coherent beams are swapped. Here the $f_{\gamma/{p,A}}$ is the photon PDF in a proton (ion), which is defined as in e.g.~\cite{Harland-Lang:2019eai}:
\begin{align}\nonumber
  x f_{\gamma/{p,A}}(x,\mu^2) &= 
  \frac{1}{2\pi \alpha(\mu^2)} \!
  \int_x^1
  \frac{dz}{z}
  \int^{\frac{\mu^2}{1-z}}_{\frac{x^2 m_{i}^2}{1-z}} 
  \frac{dQ^2}{Q^2} \alpha^2(Q^2)
  \\  \label{eq:xfgamma-phys}
  &\cdot\Bigg[\!
  \left(
    zp_{\gamma q}(z)
    + \frac{2 x ^2 m_i^2}{Q^2}
  \right)\! F_2^{p,A}(x/z,Q^2)
    -z ^2
  F_L^{p,A}\left(\frac{x}{z},Q^2\right)
  \Bigg]\,,
\end{align}
 where $m_i$ is the nucleon or ion mass, depending on the PDF under consideration. By considering the relative sizes of these photon PDFs, or more precisely the luminosity:
 \be\label{eq:lumis}
 \frac{{\rm d}\mathcal{L}}{{\rm d}m_{\gamma\gamma}^2}=   \int  {\rm d}x_1 {\rm d}x_2\, f_{\gamma/p}(x_1,\mu^2)\, f_{\gamma/A}(x_2,\mu^2) \delta(s x_1 x_2 - m_{\gamma\gamma}^2)\;,
 \ee
with $\mu^2=m_{\gamma\gamma}^2$, we can estimate the relative size of the incoherent to incoherent cross sections.
 
 \begin{figure}
\begin{center}
\includegraphics[scale=0.63]{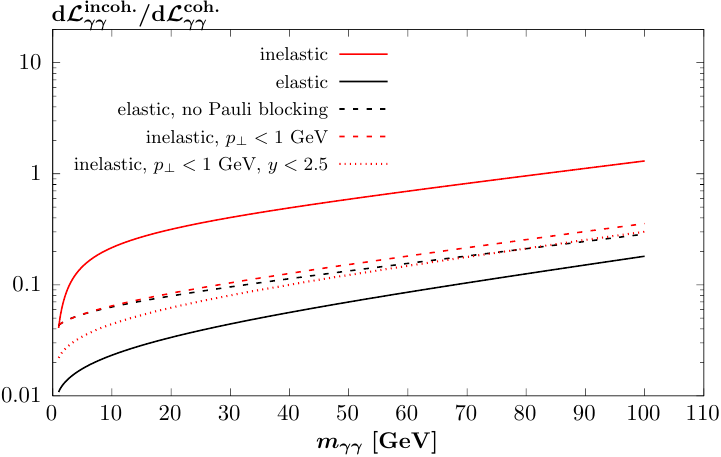}
\includegraphics[scale=0.63]{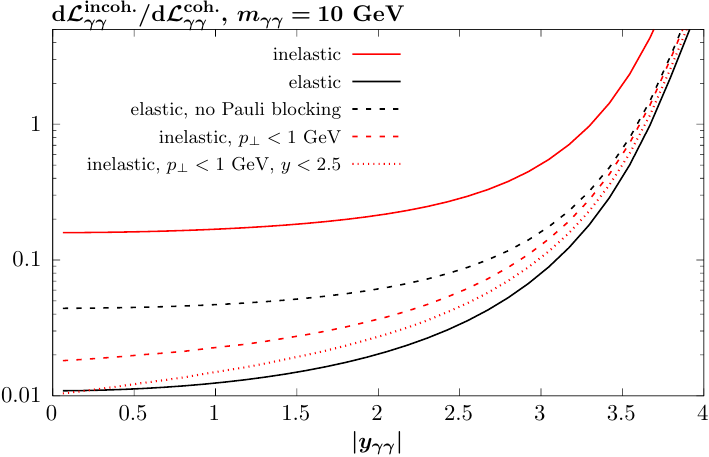}
\caption{Ratio of incoherent to coherent luminosities~\eqref{eq:lumis} as a function of (left) the diphoton invariant mass, and (right) the absolute diphoton rapidity, for $m_{\gamma\gamma} = 10$ GeV. Results are also shown in the inelastic case with additional cuts placed on the final--state quark in $q\to q\gamma$ emission generating the corresponding photon PDF, assuming LO kinematics.}
\label{fig:lumis}
\end{center}
\end{figure}

For the incoherent case, we will scale the proton (neutron) PDF contributions appropriately by $Z$ ($A-Z$), and we consistently account for the fact that either ion can interact incoherently. We will also consider the impact of applying further cuts on the final state. In particular, for UPC processes it is standard to apply a cut on the transverse momentum of centrally produced system, $p_T^X$, to be below $1-2$ GeV, in order to reduce the impact of various non--exclusive backgrounds. The incoherent case due to inelastic photon emission from the nucleons in the ion will be dominantly affected by such a cut, as here the inelastic nucleon form factors are significantly broader in $Q^2$ than the elastic and purely coherent cases. To estimate the impact of this, we can follow the approach of~\cite{Harland-Lang:2016apc} and consider modifications to the photon PDF~\eqref{eq:xfgamma-phys} for the incoherent, inelastic case. We can estimate this by considering LO kinematics for the $q\to q\gamma$ splitting that generates the inelastic photon PDF, and then applying a cut of $p_\perp < 1$ GeV on the final--state quark in such an emission. We  in particular require that
\be\label{eq:pqcut}
p^q_\perp = \left[(1-z) Q^2\right]^{1/2} < 1\, {\rm GeV}\;.
\ee
As the central system transverse momentum is dominantly given by the recoil against the inelastic side this will provide a good matching to a $p_T^X < 1$ GeV cut. Further to this, exclusivity vetos are applied in experimental analyses, by requiring that no additional activity is present within the veto region. Without going into the specific experimental configurations of any given analysis, we can provide some demonstration of this by simply requiring that the rapidity of the final--state quark, again applying LO kinematics, lies above 2.5 units for the case that $\sqrt{s_{nn}}=5.02$ TeV; see~\cite{Harland-Lang:2016apc} for a definition of the kinematics in this case.
 
Results are shown in Fig.~\ref{fig:lumis} for the ratio of the incoherent to coherent luminosities, as a function of (left) the diphoton invariant mass, and (right) the absolute diphoton rapidity, for $m_{\gamma\gamma} = 10$ GeV. Focussing first on the inelastic case, we can see that absent additional cuts this constitutes $\sim 20-100\%$ of the coherent signal, depending on the invariant mass region considered. As $m_{\gamma\gamma}$ increases, this in particular increases significantly in size, due to the larger phase space available for inelastic emission (although we note that for larger invariant masses the absolute cross section is strongly suppressed). 

This result is in line with the recent study of~\cite{Klusek-Gawenda:2025ffh} (see Fig. 5 (left), published version), where based on this it is speculated that such incoherent emission may form a sizeable component of the coherent signal for existing experimental data. However, we can see from our results that once a cut as in~\eqref{eq:pqcut} is applied, to match the signal selection in such analyses, the incoherent inelastic contribution is significantly reduced, as we would expected from the fact that a large fraction of the photon PDF will come from the higher $Q^2$ region that is removed by such cut. The result is now at the $\sim 2 - 10\%$ level. For the event selection of most existing data, for which lower values of $m_{\gamma\gamma}$ dominantly contribute, the leads us to expect a percent level contribution. Applying the rapidity veto cut on top of this further reduces this, as we can see from the figure.

Turning to the elastic case, before Pauli blocking effects~\eqref{eq:pauli} are applied this provides a comparable contribution to the inelastic case, after imposing suitable cuts. However, Pauli blocking reduces this by a factor of $\sim 3-5$. The final contribution is again at the percent level. For the luminosity ratios vs. diphoton rapidity, similar conclusions hold. We can see that at very forward rapidity the incoherent contribution begins to dominate, due to the sharply falling coherent flux with $x$, but these are in a region where the absolute cross section is strongly suppressed, and are also outside current experimental selections.

\begin{figure}
\begin{center}
\includegraphics[scale=0.63]{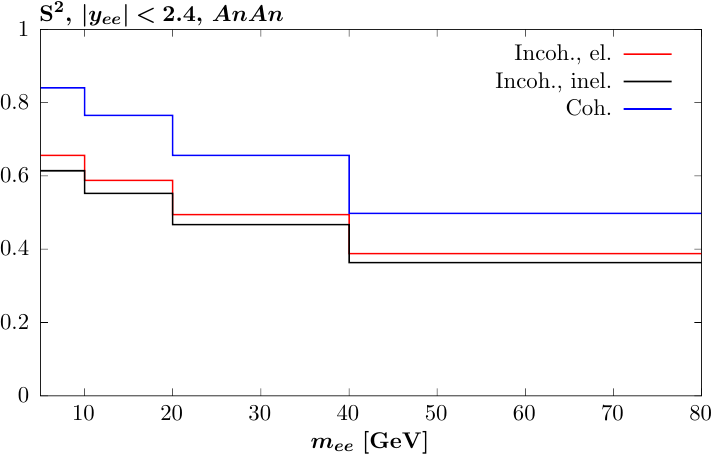}
\caption{Predicted survival factors for coherent and incoherent electron pair production, within the ATLAS~\cite{ATLAS:2022srr} event selection, for $|y_{ee}|<2.4$ and as a function of the dielectron invariant mass.}
\label{fig:s2}
\end{center}
\end{figure}

Finally, we note that in addition to the above effects, the survival factor probability of no addition ion--ion interactions must be accounted for, as discussed in the previous section. This is shown in Fig.~\ref{fig:s2} for the event selection of the ATLAS measurement~\cite{ATLAS:2022srr} of electron pair production in PbPb UPCs, which we will consider again in the following section. We can see that this provides some further suppression in the incoherent cross sections, that is somewhat more significant than in the purely coherent case. This is as we may expect given that even for nucleon--ion impact parameters that are themselves larger than the ion radius, for some ion--ion geometries there may nonetheless be ion overlap (see Fig.~\ref{fig:ionion}).

\subsection{Comparison to ATLAS data on dielectron production}\label{sec:atlasdiel}

In this section we present predictions for electron pair production, and comparisons to the ATLAS measurement~\cite{ATLAS:2022srr} of this process. The full event selection is as described in this paper, but of particular relevance is the cut, $p_\perp^{ee} < 2$ GeV, on the dielectron transverse momentum, which as discussed about tends to reduce the incoherent cross section rather significantly.  In this analysis, a template fit is performed to the dielecton acoplanarity, $\alpha = 1-\Delta\phi_{ee}/\pi$, with the coherent signal distribution modelled by the \texttt{Starlight} MC~\cite{Klein:2016yzr} and incoherent distribution modelled using single--dissociative events in $pp$ collisions from \texttt{SuperChic}~\cite{Harland-Lang:2020veo} (the only available possibility at the time), in both cases interfaced to \texttt{Pythia}~\cite{Sjostrand:2014zea} to account for FSR effects. In the latter case, as it is the same inelastic proton structure functions that enter the calculation of incoherent inelastic production presented here, we expect the acoplanarity distribution calculated in this way to closely match the results here; in particular, while the elastic beam is a proton in the ATLAS sample, rather than an ion, and hence has a broader $Q^2$ distribution, the dominant contribution to the dielectron acoplanarity in the tail region is due to the inelastic beam.

\begin{figure}
\begin{center}
\includegraphics[scale=0.63]{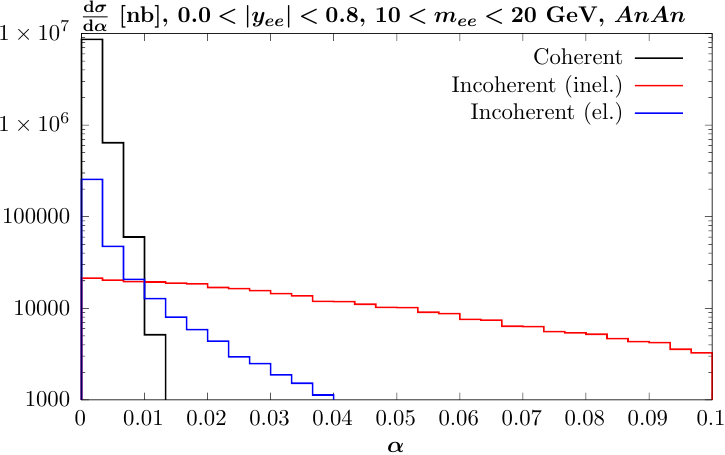}
\includegraphics[scale=0.63]{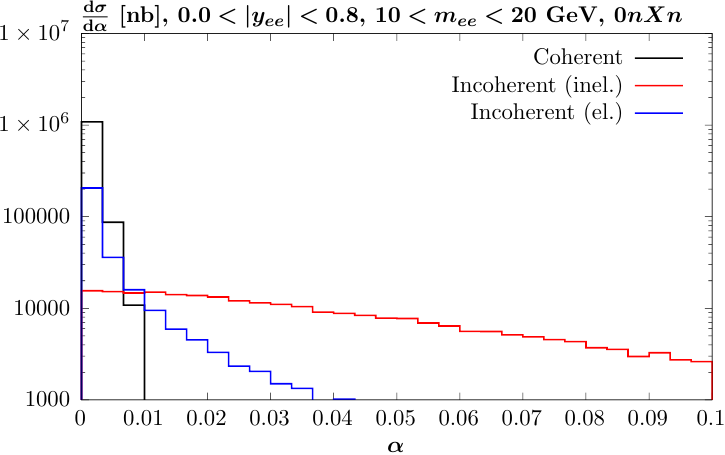}
\caption{Predicted coherent and incoherent electron pair production cross sections, differential with respect to the dielectron acoplanarity, within the ATLAS~\cite{ATLAS:2022srr} event selection, for $|y_{ee}|<0.8$. Results are given for different ZDC selections, as indicated.}
\label{fig:ATdiel_aco}
\end{center}
\end{figure}

Predictions for these acoplanarity distributions are shown in Fig.~\ref{fig:ATdiel_aco}, with the dielectron rapidity chosen to match the ATLAS results (Fig. 3 of~\cite{ATLAS:2022srr}). As these are not provided corrected to the cross section level in the ATLAS analysis we do not present a direct comparison, but broadly speaking the trend observed in the data, and expected on general grounds, is present. In particular, the data are dominantly composed of a coherent signal that is  sharply peaked at low acoplanarity, and a much broader tail due to the incoherent inelastic contribution. The incoherent elastic contribution has a broader acoplanarity distribution than the coherent signal, but this nonetheless falls significantly faster than the inelastic case, as we would expect. Due to the difference between this and the inelastic case, this contribution may therefore not be in general accounted for in the two component template fit performed by ATLAS,  although as we will see below it is expected to be subdominant.

Results are also shown for the $0nXn$ ZDC selection. This is accounted for following the approach of~\cite{Harland-Lang:2023ohq}, although some care is needed in the interpretation. In particular, for coherent production we simply include the probability for the $0nXn$ (which we define to include $Xn0n$ as well) final--state, i.e. with additional ion excitation of one beam and without it for the other. However, for incoherent production (both elastic and inelastic) the incoherent ion will breakup and will therefore automatically lie in the $Xn$ category. Therefore, for the $0nXn$ category, we should multiply the incoherent cross sections by a $0nAn$ excitation probability, if beam 2 interacts inelastically, where in this case we do not include the corresponding $An0n$ possibility (which would correspond in part to the  a $XnXn$ final--state), and vice versa if beam 1 interacts inelastically. For the $XnXn$ case, this is calculated in the usual way for  coherent production, while for incoherent production it is by definition given by the remaining cross section that is not $Xn0n$. Broadly, the result of this is that the $0nXn$ requirement tends to reduce the coherent cross section more significantly than the incoherent, as is observed in  Fig.~\ref{fig:ATdiel_aco} (right). 

Turning now to a more direct comparison to the data, incoherent to coherent cross section fractions are presented in~\cite{ATLAS:2022srr}, extracted using the template fitting approach described above, and for certain requirements on the dielectron rapidity and invariant mass. In the main published analysis, these are in particular given for the case that an additional $|y_{ee}|<0.8$, $10 < m_{ee}<20$ GeV cut is placed in the dielectron system. The results are given in Table~\ref{tab:fxx} for the three relevant ZDC selections considered in the analysis. While the $0n0n$ case is also considered by ATLAS, this is predicted to have no incoherent contribution, and indeed the ATLAS data are consistent with this; hence we do not consider this case further in what follows. We can see that broadly the predicted fractions match the data rather well. We will discuss the theoretical uncertainties further below but note that a $\sim 20\%$ uncertainty (at least) is certainly expected. Hence within these (and the experimental) uncertainties the agreement is very good.

\begin{table}
\begin{center}
\begin{tabular}{|c|c|c|c|}
\hline
&AnAn  & 0nXn& XnXn\\
 \hline
 ATLAS~\cite{ATLAS:2022srr},$ |y_{ee}|<0.8$, $10 < m_{ee}<20$ GeV&$0.043\pm0.0026$&$0.099\pm0.006$&$0.13\pm0.01$\\
  \hline
Central prediction&0.039&0.12&0.17\\
\hline
\end{tabular}
\end{center}
\caption{Predicted and measured incoherent, inelastic electron pair production fractional contributions to the purely coherent case, within the ATLAS~\cite{ATLAS:2022srr} event selection, for $|y_{ee}|<0.8$, $10 < m_{ee}<20$ GeV.}  \label{tab:fxx}
\end{table}

 \begin{figure}
\begin{center}
\includegraphics[scale=0.63]{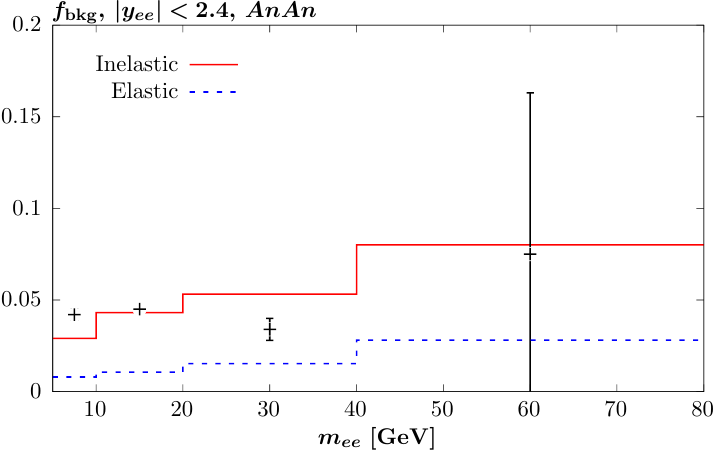}
\includegraphics[scale=0.63]{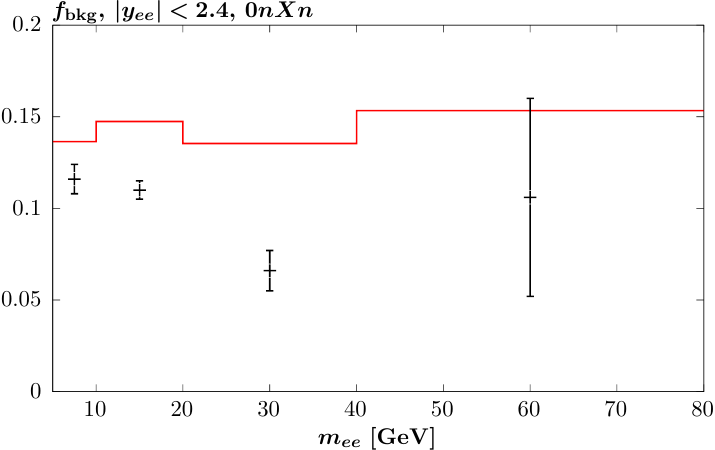}
\includegraphics[scale=0.63]{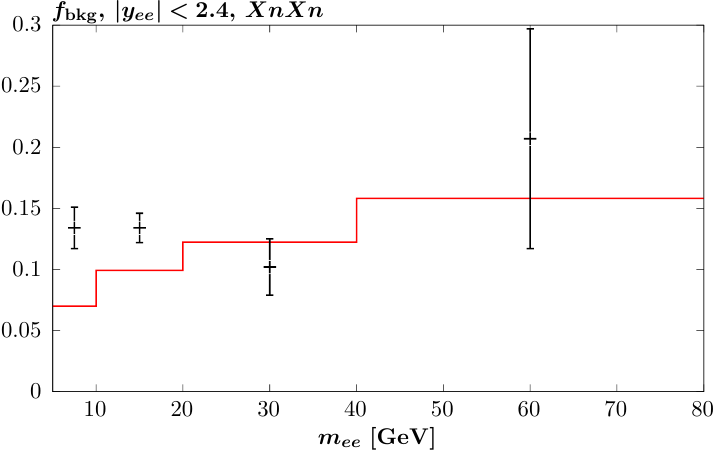}
\caption{Predicted incoherent, inelastic electron pair production fractional contributions to the purely coherent case, within the ATLAS~\cite{ATLAS:2022srr} event selection, for $|y_{ee}|<2.4$ and as a function of the dielectron invariant mass. The measured (unpublished) ATLAS values  taken from~\cite{Ogrodnik:2022srz}, and results are given for different ZDC selections, as indicated. For the $AnAn$ case the elastic incoherent contribution is also shown for comparison, but we emphasise that the ATLAS data correspond dominantly to the purely inelastic contribution, and hence this should not be included when comparing to these data.}
\label{fig:ATdiel_y0024}
\end{center}
\end{figure}

 \begin{figure}
\begin{center}
\includegraphics[scale=0.63]{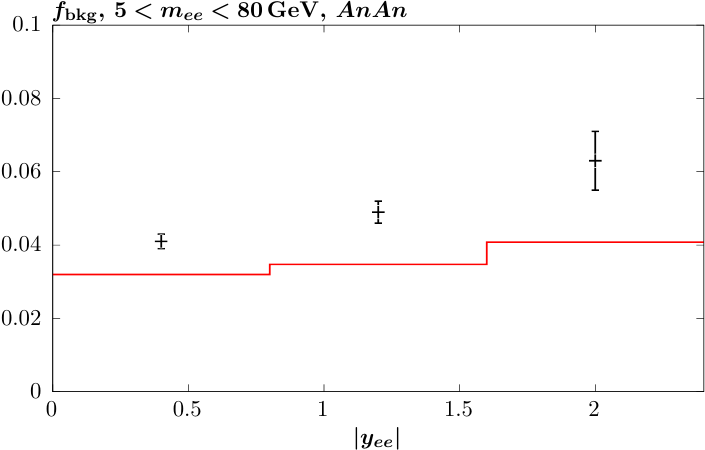}
\includegraphics[scale=0.63]{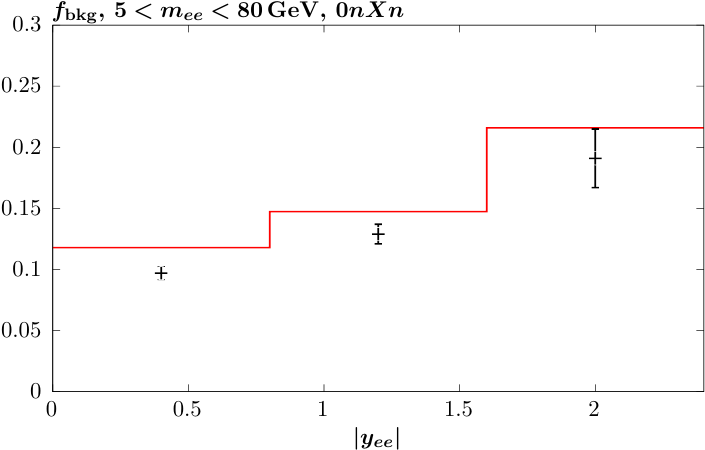}
\caption{Predicted incoherent, inelastic electron pair production fractional contributions to the purely coherent case, within the ATLAS~\cite{ATLAS:2022srr} event selection, for $5 < m_{ee} <80$ GeV and as a function of the magnitude of the dielectron  rapidity. The measured ATLAS values, taken from~\cite{Ogrodnik:2022srz}, are also shown, and results are given for different ZDC selections, as indicated.}
\label{fig:ATdiel_m580}
\end{center}
\end{figure}

Further comparisons to the unpublished results presented in~\cite{Ogrodnik:2022srz} are shown in Figs.~\ref{fig:ATdiel_y0024} and~\ref{fig:ATdiel_m580}, as a function of the dielectron invariant mass and rapidity. Broadly speaking, the agreement is rather good, although in more detail there are regions, e.g. in the lowest invariant mass bin and for the $0nXn$ ZDC selection, where the agreement is somewhat less good. Nonetheless, this can give us some broad confidence in the overall results. In particular, and looking ahead to Section~\ref{sec:lbylgam}, this level of agreement validates  the predictions for the incoherent background in the case of LybL scattering to a reasonable level of precision. 

In terms of the uncertainties on the predictions, a straightforward source of uncertainty on the inelastic contribution comes from the PDF uncertainty on the parton--level process. The PDF uncertainty due to the initial--state quark/antiquarks can be calculated using the standard \texttt{MSHT20qed\_nnlo} PDFs~\cite{Cridge:2021pxm} eigenvectors, and is found to be rather small, $\sim 1\%$. There is in addition an uncertainty due to the modelling of the lower $Q^2,W^2$ region, but this is lower still. In broad, terms, we expect this source of uncertainty to follow that of the photon PDF itself in~\cite{Cridge:2021pxm}, i.e. indeed be at the percent level. The above PDF set is however that of a free proton, and hence no account of nuclear modifications to the proton PDF has been accounted for here. To evaluate this, we have investigated the impact of changing from the $p$ to $Pb$ PDFs using the nNNPDF3.0 NLO PDF sets~\cite{AbdulKhalek:2022fyi}. Using the Pb set tends to increase the predicted cross section somewhat, by $\sim 2-5\%$, depending on the kinematic region considered. Further to this, there is some uncertainty due to the modelling of the survival factor, although as discussed in~\cite{Harland-Lang:2021ysd} this is expected to be very small, and moreover it will to some extent cancel in the cross section ratios considered here. 

The naive uncertainty is therefore at the $\sim 5\%$ level, and is hence rather small. However, this does not account for what is most likely to be the dominant uncertainty source, namely the impact of the experimental exclusivity requirement on the data. We do not apply any such veto in the above comparisons, however we have seen from Fig.~\ref{fig:lumis} that imposing a simple cut on the final--state quark assuming LO $q\to q\gamma$ kinematic in the inelastic process leads to a $\sim 20\%$ reduction in the expected cross section. A more precise calculation would  a full account of the particle multiplicity distribution of the ion dissociation system, which is beyond the scope of the current study. Nonetheless, conservatively we can take an uncertainty of that order here, although we note that this effect will only act to reduce the predicted cross sections, which can therefore be viewed as upper bounds from this point of view.

In the elastic case, the dominant source of uncertainty is due to the modelling of Pauli blocking effect as in~\eqref{eq:pauli}. We have seen from  Fig.~\ref{fig:lumis} that this reduces the predicted contribution by a significant factor of $\sim 3-5$. The physical origin of this effect is well understood, and accounting for it is a key feature of modelling nuclear scattering experiments, e.g. in the case of electron~\cite{Bosted:2012qc} and neutrino scattering off nuclear targets (see~\cite{SajjadAthar:2022pjt} for a recent summary).
However there is certainly some uncertainty in the precise implementation of this effect. To asses this, we investigate the effect of reducing the Fermi momentum used in~\eqref{eq:pauli1} from 250 to 200 MeV. While this represents a rather aggressive level of reduction that is not necessarily favoured by data or models of nuclear structure, it will provide some indication of the level of model dependence. We find that this tends to increase the elastic cross section by $\sim 10-20\%$, and so we can take this as a representative estimate for the corresponding uncertainty.

\section{Light by Light scattering}\label{sec:lbylgam}

\subsection{Signal--like contribution}\label{sec:signallike}

 \begin{figure}
\begin{center}
\includegraphics[scale=0.63]{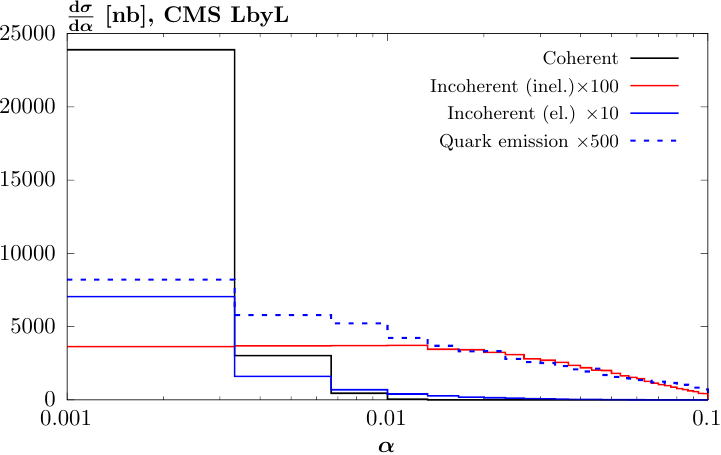}
\caption{Predicted coherent and incoherent light--by--light scattering cross sections, differential with respect to the diphoton acoplanarity, within the CMS~\cite{CMS:2024bnt} event selection. The incoherent contributions are scaled as indicated in the legend for ease of comparison. The incoherent `quark emission' case, with a $\eta_q > \pm 3$ cut imposed, depending on the relevant incoherent beam, is also shown; this is discussed in Section~\ref{sec:qe}.}
\label{fig:CMSlbyl_aco}
\end{center}
\end{figure}

We now consider the case of LybL scattering, in particular having in mind the large acoplanarity backgrounds that are observed (and subtracted, using a template fit) in the ATLAS~\cite{ATLAS:2020hii} and CMS~\cite{CMS:2024bnt} measurements of this process. We begin by considering the `signal--like' incoherent contribution to this process, i.e. as shown in Fig.~\ref{fig:Feyn_incoh} (right), while in the following section we will discuss additional contributions that should in general be accounted for here.

The predicted acoplanarity distributions for coherent and incoherent production are shown in Fig.~\ref{fig:CMSlbyl_aco} for the CMS event selection. For the ATLAS selection, the corresponding distributions are rather similar, and so are omitted for brevity. The incoherent contributions, which as we will see explicitly below enter at the percent level, are scaled by a factor of 10 (100) in the elastic (inelastic) cases for ease of comparison. This result is to be compared with Fig.~5 of~\cite{CMS:2024bnt}, where the corresponding measurement is shown. We can see that the shape of the coherent contribution is clearly peaked towards low acoplanarity values, as expected, while the incoherent inelastic leads to precisely the sort of tail that is observed in the data. However, it is only after scaling this inelastic contribution by a factor of 100 that the size of this is seen to roughly match that observed in the data. The shape of the elastic contribution  is rather more similar to the signal case, but is again significantly smaller than it. Finally, the contribution from the `quark emission' process that we will discuss further in Section~\ref{sec:qe} is shown for comparison.

\begin{table}
\begin{center}
\begin{tabular}{|c|c|c|}
\hline
& Incoherent, elastic & Incoherent, inelastic\\
 \hline
 ATLAS~\cite{ATLAS:2020hii} selection&0.9&2.1\\
  \hline
 ATLAS~\cite{ATLAS:2020hii} selection, $\alpha < 0.01$&0.5&0.4 \\
  \hline
 CMS~\cite{CMS:2024bnt} selection&0.9&2.3\\
   \hline
 CMS~\cite{CMS:2024bnt} selection, $\alpha < 0.01$&0.4&0.4\\
\hline
\end{tabular}
\end{center}
\caption{Predicted percentage `signal--like' contribution from incoherent production to the coherent light--by--light scattering, within the ATLAS~\cite{ATLAS:2020hii} and CMS~\cite{CMS:2024bnt}  event selections. Results with and without the acoplanarity cut $\alpha< 0.01$, are shown for comparison, although  for both measurements this cut is applied.}  \label{tab:lbyl}
\end{table}

In more detail, in Table~\ref{tab:lbyl} we show the predicted percentage contribution from incoherent production to  coherent LybL scattering, within the ATLAS~\cite{ATLAS:2020hii} and CMS~\cite{CMS:2024bnt}  event selections. We can see that for both elastic and inelastic production these are at the $\sim 0.5$\% level after  the acoplanarity cut $\alpha< 0.01$ is imposed, as it is for both measurements.  We note that if the acoplanarity cut is not imposed the inelastic contribution is instead $\sim 2\%$, and if we in addition increase the upper cut on the diphoton transverse momentum from 1 to 2 GeV to match that imposed in the ATLAS dielectron analysis~\cite{ATLAS:2022srr} then this increases further to $\sim 4\%$, i.e. in line with the prediction in Table~\ref{tab:fxx}, as we would expect given the similar kinematic regions and the fact that the relative incoherent contribution from this type of diagram should be relatively process independent. 

Once the tighter $p_\perp$ and acoplanarity cuts are imposed, on the other hand, incoherent production is  expected to provide a very minor contribution in the signal region, and therefore cannot explain the size of the larger acoplanarity background observed in both analysis. In particular, although in~\cite{ATLAS:2020hii,CMS:2024bnt}  precise cross section extractions are not given, the number of background events is quoted, and gives a good indication of the measured backgrounds. In both of these analyses a template fit is performed, with the shape of the high acoplanarity background tail modelled using the~\texttt{SuperChic} prediction~\cite{Harland-Lang:2018iur} for QCD--initiated diphoton production. The normalization is then fitted to the data, along with the coherent signal. As we will discuss further in Section~\ref{sec:QCD}, while this production mechanism is different, this results in a similar (broad) shape to the acoplanarity distribution to the PI incoherent inelastic contribution, and hence such extractions can be treated relatively agnostically with respect to the precise production mechanism.

It is found in the  ATLAS case that the ratio of background to signal events is $12/45 \sim 0.27$, while in the CMS case it is $10/13 \sim 0.77$.  Given the significant statistical uncertainty on the latter result these are roughly consistent with each other. Therefore, broadly speaking these suggest a ratio of background to signal of $\sim 0.5$, which is clearly significantly higher, by roughly two orders of magnitude, than the predictions in Table~\ref{tab:lbyl}. In other words, these cannot be the dominant background contribution. We recall in particular that the theoretical uncertainty on these predictions is expected to be  $O(10\%)$, while the good agreement with the ATLAS data on dielectron production lends further evidence to the fact that this background is accurately modelled. 

 In the elastic case,  we can see again from Fig.~\ref{fig:CMSlbyl_aco} that the predicted acoplanarity distribution is significantly narrower than the inelastic component, and more in line with the coherent signal. Therefore such a contribution will not contribute significantly to the observed backgrounds. On the other hand, it will also not be effectively subtracted by the template fitting procedure, and hence should be explicitly accounted for in future analyses, even if as we can see the expected contribution is small.

\subsection{Quark emission diagrams}\label{sec:qe}

\begin{figure}
\begin{center}
\includegraphics[scale=0.7]{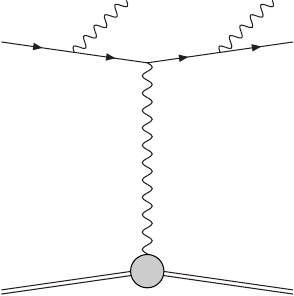}
\caption{Representative `quark emission' Feynman diagram for incoherent contribution to light--by--light scattering signal in AA ultraperipheral collisions. }
\label{fig:lbyl_pscat}
\end{center}
\end{figure}

Considering again the Feynman diagram in Fig.~\ref{fig:Feyn_incoh} in the inelastic case, namely where the upper line is interpreted as quark/antiquark parton within a constituent nucleon in the ion, it is clear that this is not the only contributing diagram. There are in particular various diagrams where either one or both photons are emitted by directly by the upper quark line. This is shown in Fig.~\ref{fig:lbyl_pscat}, where the other diagrams with the photons  emitted in different configurations by the upper quark lines are implicit. We note that there are in general various possible sources of collinear singularities in these diagrams, but these are not present in the phase space region relevant here. In particular, the final--state collinear quark--photon singularity is removed by the requirement that the photons are produced centrally and with sufficient $p_\perp$, while the outgoing quark line must lie outside the central region, or more precisely the exclusivity veto region, which can extend further again. The initial--state  singularity, due to a collinear 
splitting from the coherent photon, is similarly removed by the exclusivity requirement that the final--state quark lies in the opposite hemisphere to the direction of this photon.

This class of diagram is particularly relevant as we can see by comparison of the two cases that it actually enters at a lower order in $\alpha$. In particular, while the `signal--like' incoherent production cross section is $O(\alpha^5)$, the quark emission cross section is $O(\alpha^3)$, and hence naively should be enhanced by a significant fact of $1/\alpha^2 \sim 10^4$, which given the signal--like incoherent contribution enters at the percent level of the coherent, might suggest that this is even larger (by a factor of $\sim 10^2$) then the purely coherent signal. We note that this difference in the order in $\alpha$ also clearly demonstrates that there is no question of gauge invariance from treating the diagrams individually, although in principle they should be added at the amplitude level, with interference included; we will comment further on this below.

\begin{figure}
\begin{center}
\includegraphics[scale=0.56]{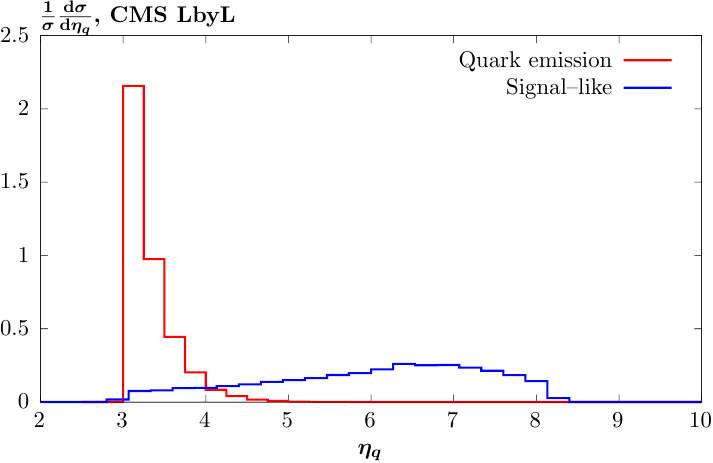}
\caption{Normalized distribution with respect to the pseudorapidity, $\eta_q$, of the outgoing quark line, in the lab frame, for `quark emission'  and `signal--like' incoherent contributions to light--by--light scattering signal in AA ultraperipheral collisions. The case where the incoherently interacting ion is in the positive $z$ direction is considered for concreteness, and the final--state photons lie within the CMS~\cite{CMS:2024bnt} LbyL scattering fiducial region, while a cut of $\eta_q > 3$ is imposed.}
\label{fig:lbyl_pscat_etaq}
\end{center}
\end{figure}

However, there are various factors that will counteract this enhancement. First, it is known from basic MHV considerations that the $\gamma(p_a) q(p_b) \to \gamma(p_1)\gamma(p_2) q(p_b')$ amplitude~\cite{Ozeren:2005mp} is proportional to the spinor production of the incoming and outgoing quark momenta
\be\label{eq:mhv}
\mathcal{M}(\gamma(p_a) q(p_b) \to \gamma(p_1)\gamma(p_2) q(p_b')) \propto \langle p_b p_b' \rangle\;.
\ee
This implies (combined with the fact that the rest of the amplitude is perfectly regular in this limit) that the scattering amplitude in fact vanishes in the limit of forward quark scattering, with the outgoing quark momentum $p_b' \propto p_b$. As this is precisely the limiting case of the phase space region of interest, namely where the outgoing quark travels in the forward direction and does not enter the rapidity veto region, we can expect this to suppress the corresponding cross section. Indeed, an explicit check of the e.g. outgoing quark pseudorapidity distribution demonstrates that the contribution from the forward region is strongly suppressed relative to the standard `signal--like' case, with a preference for the phase space region towards the edge of the veto region. We can see this in Fig.~\ref{fig:lbyl_pscat_etaq}, which shows the normalized distribution with respect to the outgoing quark pseudorapidity, $\eta_q$, for both of these contributions. This is defined in the lab frame, which is distinct from the nucleon--ion rest frame in which the incoherent cross section is typically calculated. The difference in the two cases is clear, although we note that this is also driven by the $t$--channel enhancement from the photon propagator in the signal--like case that is not present for the quark emission process.

In addition to this, we can see that the signal--like diagram in general receives a $t$--channel $\sim 1/Q^2$ enhancement from the upper photon propagator\footnote{We note that as usual one factor of $1/Q^2$ is effectively cancelled by the numerator of the upper $qq\gamma$ vertex.}, which in the quark emission diagrams is replaced by a factor of order $\sim 1/M_{\gamma q}^2$, where $M_{\gamma q}$ is the invariant mass of the initial--state quark--photon system. For the inelastic incoherent case we are considering the average photon $Q^2$ is $O({\rm GeV}^2)$ and is significantly larger than in the case of elastic incoherent, and certainly purely coherent, production. Nonetheless this will clearly lead to a relative suppression in comparison to the signal--like incoherent case, which increases with $M_{\gamma q}$ and hence the diphoton invariant mass $M_{\gamma\gamma}$. Very roughly speaking, if we assume that $\left \langle Q^2 \right \rangle  = 1\, {\rm GeV}^2$ and $M_{\gamma q} \sim M_{\gamma\gamma}$ then for LybL scattering in the ATLAS and CMS event selection this will lead to a suppression of
\be\nonumber
\frac{1 {\rm GeV}^2}{M_{\gamma\gamma}^2} \sim \frac{1}{25}\;,
\ee
i.e. 1-2 orders of magnitude, and in reality given $M_{\gamma q} > M_{\gamma\gamma}$, closer to 2, although clearly the precise level of suppression will depend on the specifics of the processes and their kinematics. Nonetheless, we can broadly expect this to reduce the overall enhancement of the quark emission process with respect to the signal--like from $\sim 10^4$ to $\sim 10^2$, and further at higher masses.

Moreover, we note that the LybL scattering, $\gamma\gamma \to \gamma\gamma$, cross section itself receives some further numerical enhancement from the summation over  the different particles in the intermediate loop. From~\cite{Bern:2001dg} the (dominant) lepton loop contribution to the LO amplitude for e.g. all positive photon helicities is given  by
\be
\mathcal{M}_{++++} = 8 \alpha^2\;,
\ee
in the massless lepton limit. Hence summing over the three lepton species we have an enhancement of $\sim (8*3)^2 \sim 600$. Clearly the precise level of enhancement will depend on the precise structure of the corresponding quark emission diagrams, but we note that given the photons are in this case purely emitted from the quark line, this will lead to a further $\sim 1/e_q^4$ suppression with respect to the signal--like case, i.e. by roughly an order of magnitude. 

Finally, we recall that there are additional cuts placed on the diphoton system, namely to restrict the transverse momentum to be $p_\perp^{\gamma\gamma}< 1$ GeV as well as on the acoplanarity. Given the individual transverse momenta are peaked at 2.5 GeV, i.e. not significantly higher than this cut, this will not necessarily reduce the cross section dramatically, in particular having in mind that we are considering the suppression relative to the incoherent signal--like contribution, which is also suppressed by such a cut. Nonetheless this will lead to further suppression, which will certainly become more relevant for higher photon transverse momenta.

Putting the above effects together, we may expect this contribution to enter at something like the same level as the purely signal--like incoherent contribution, but not to be significantly enhanced with respect to it. To verify this, we have implemented this process following the approach of~\cite{Bailey:2022wqy}. Namely, this type of diagram can be evaluated using the general expression \eqref{eq:sighhf} but making the substitution
\be\label{eq:rhorep}
\rho_{1}^{\mu\mu'}\rho_{2}^{\nu\nu'} M^*_{\mu'\nu'}M_{\mu\nu} \to \frac{Q_1^2}{4\pi \alpha(Q_1^2)} \int  \frac{{\rm d}M_1^2}{Q_1^2} \, \rho_{2}^{\mu\mu'} \sigma_{\mu \mu'}^{1}\;,
\ee
where the integration is as usual performed simultaneously with the other phase space integrals, and similarly if beam 2 instead interacts incoherently. At LO we have
\be\label{eq:siggq}
\sigma_{\mu \mu'}^{i} = \sum_{j=q,\overline{q}}  f_j(x_{B,i},\mu_F^2)  \left \langle A_{\mu}^i A_{\mu'}^{i*} \right\rangle \;,
\ee
where $A_\mu^i$ is the corresponding $\gamma^* + q \to \gamma \gamma + q$ amplitude as in Fig.~\ref{fig:lbyl_pscat}, with a collinear initial--state quark/anti--quark from beam $i$, carrying proton momentum fraction $x_{B,i}$. Further details of the precise implementation of this, and the relation to the standard collinear result are given in Appendix A 
 of~\cite{Bailey:2022wqy}. These amplitudes are calculated from the standalone output  of \texttt{MadGraph5\_aMC@NLO}~\cite{Alwall:2014hca,Frederix:2018nkq}.
 
As discussed above, the singularity structure of this process requires that we impose an exclusivity veto on the final--state quark, in order to remove the collinear quark--photon regions. We do this by simply imposing that the final--state quark pseudorapidity lies above some cut that is outside the central region where the photons are present. We in particular consider $\eta_q > 3$ -- 4.5 for the case that beam 1 (defined as travelling in the positive $z$ direction) interacts incoherently, and similarly in the negative rapidity region if it is beam 2. The lower limit is almost certainly too loose to match the exclusivity vetos imposed experimentally, but nonetheless gives some estimate. For simplicity, we neglect the impact of the ion--ion survival factor, though this may in general further suppress the relative contribution from the quark emission diagrams. We find that the predicted cross section is rather strongly sensitive to the precise cut, but even for the looser $\eta_q > 3$ case, the predicted cross section is a factor of $\sim 3$ lower than the signal--like contribution within the ATLAS or CMS fiducial regions. For more stringent cuts, this reduces even further. 

Therefore, we can safely ignore this type of contribution, and indeed any interference it has with the signal--like case. More broadly, there are of course various 1--loop corrections to the quark emission diagrams that would then enter at the some order in $\alpha$ as the signal--like one, but these will clearly be further suppressed, and can be neglected. We note that in the case of lepton pair production, there is a similar topology of diagram, where the lepton pair is emitted from the quark line directly. However, as discussed in Section 3 of~\cite{Bailey:2022wqy} these enter at the percent level in general in comparison to the signal--like incoherent case and hence for current purposes we can safely neglect this.

\subsection{Exclusive Two-Photon Photoproduction: Brief Remarks}\label{sec:photoproduction}

Having considered the leading--order inelastic contribution to the case that the two photons are emitted directly from the interacting nucleon, as per Fig.~\ref{fig:lbyl_pscat}, it is worth commenting on the fact that such a process can also proceed purely elastically, that is with the outgoing nucleon (but not ion) remaining intact. Indeed, this has been discussed in detail in~\cite{Pedrak:2017cpp,Grocholski:2022rqj,Grocholski:2021man} in the context of lepton--nucleon scattering. The key observation here is that the production amplitude is purely imaginary, and proportional to the valence generalized quark distributions taken at the border values of $\xi = \pm x$ (i.e. precisely in the skewed regime relevant to gluon--initiated CEP processes~\cite{Harland-Lang:2014lxa}, in that case for the generalized gluon PDF). 

As an aside, we note that the cancellation of the real amplitude in fact follows from the same observation made above, see~\eqref{eq:mhv}, given the amplitude precisely corresponds to such forward kinematics. However, from the point of view of phenomenology the relevant factor is that the C--parity of the overall proceeds leads to only the generalized valence quark distributions contributing, which will therefore strongly suppress the contribution from this process, in particular for the intermediate $x$ values of relevance here. Therefore it is not expected to be a relevant source of background to the LbyL scattering signal. It remains possible however that this $t$--channel quark exchange topology may nonetheless play a role if we move away from the purely exclusive region, although in general this will be suppressed by the cut on the diphoton $p_\perp$; a detailed analysis of these issues is left to future work.

\subsection{QCD--initiated production revisited}\label{sec:QCD}

Having considered the various possible purely QED contributions for the higher acoplanarity background to LybL scattering, we have seen that these are not expected to provide anything beyond a percent--level correction in the signal (low acoplanarity) region. However, as discussed above, this is in contrast to the data--driven determinations by both ATLAS~\cite{ATLAS:2020hii} and CMS~\cite{CMS:2024bnt}, which find contributions that are at least an order of magnitude higher than this. In both of these analyses, the  shape of the high acoplanarity background tail is modelled using the~\texttt{SuperChic} prediction~\cite{Harland-Lang:2018iur} for QCD--initiated diphoton production. However the normalization is allowed to be free in the fit, and is also found to be significantly higher than this QCD--initiated contribution predicted in~\cite{Harland-Lang:2018iur}, by roughly an order of magnitude or more.

With this in mind, we briefly revisit the case of QCD--initiated production here, and discuss the extent to which such a larger normalization could be consistent with expectations. We recall that this process proceeds according to the `Durham' perturbative  QCD model for exclusive QCD--initiated production (see e.g.~\cite{Harland-Lang:2014lxa} for a review), namely the underlying process is due $gg\to \gamma\gamma$ scattering, i.e. exactly as in Fig.~\ref{fig:Feyn_incoh} (left) but with the initial--state photons replaced by gluons, while an additional soft `screening' gluon must then be exchanged between the interacting nucleons to ensure a colour--singlet exchange. 

As discussed in detail in~\cite{Harland-Lang:2018iur}, when generalising the proton--proton case to that of heavy ion collisions, we can consider two possibilities, namely coherent and incoherent production. In the former case, the ion acts as a coherent source of gluons and remains intact afterwards, while in the latter we simply consider the case that the individual nucleons can act incoherently as individual sources of gluons, with the ion breaking up, in both cases in the appropriate colour--singlet configuration; these mechanisms are fully analogous to the cases of coherent and incoherent PI production discussed above. 

In~\cite{Harland-Lang:2018iur} the cases where both ions interacted either coherently or incoherently  were considered, and so for completeness we extend this here to consider the `mixed' case where one ion interacts coherently and one interacts incoherently. We have also improved the modelling of the incoherent case, making use of the approach discussed previously for the case of PI production. In particular, in~\cite{Harland-Lang:2018iur} the ion--ion survival factor was accounted for by multiplying the nucleon--nucleon cross section by
 \be\label{eq:s2incoh}
 \left \langle S^2_{\rm incoh} \right \rangle  = \frac{\int {\rm d}^2 b_{1\perp} {\rm d}^2 b_{2\perp} T_n(b_{1\perp}) T_n(b_{2\perp}) e^{-\Omega_{A_1 A_2}(b_{1\perp}-b_{2\perp})}}{\int {\rm d}^2 b_{1\perp} {\rm d}^2 b_{2\perp} T_n(b_{1\perp}) T_n(b_{2\perp}) }\;,
 \ee
 where $b_{i\perp}$ are as usual the impact parameters between the ion centres and the interacting nucleons. In other words, this approximates the result by neglecting the range of the underlying nucleon--nucleon interaction. We can however readily account for this by generalising the approach described in Section~\ref{sec:surv}. Namely~\eqref{eq:gaminelel} becomes
 \be\label{eq:gaminelel_gen}
\Gamma^{\rm QCD, incoh}_{A_1A_2}(s,b_\perp)=   \int {\rm d}^2 \tilde{b}_{1\perp} {\rm d}^2 \tilde{b}_{2\perp} T_A(\tilde{b}_{1\perp}) T_A(\tilde{b}_{2\perp}) \Gamma_{AA}(s,b_\perp') \;,
\ee
where $ {\mathbf b'}_\perp = {\mathbf b}_\perp-\tilde{{\mathbf b}}_{1\perp}-\tilde{{\mathbf b}}_{2\perp}$ and then the cross section can be calculated in the same manner. The previous approach will tend to underestimate the cross section somewhat, although the effect is rather mild, at the $\sim 10\%$ level. The mixed case is calculated in a similar fashion. 

Given the sizeable large acoplanarity background observed in the ATLAS and CMS data, we also, in order to be as aggressive as is reasonably possible in terms of the size of the predicted incoherent cross section, now omit the additional $pp$ survival factor in the calculation. That is, the above expression is assumed to capture the totality of the soft survival effects. As discussed in~\cite{Harland-Lang:2018iur}, there are reasons to believe that this will overestimate the expected cross section. In particular, as the exclusive production process occurs at the ion periphery, the  nucleon density is low and hence the average number of nucleon–nucleon interactions contained in the above expression can be below one. Nonetheless, this can serve to provide an upper bound on the cross section, with the question of accounting for the data kept in mind. Finally, we have corrected the implementation of the coherent production cross section used in~\cite{Harland-Lang:2018iur}, where the Pomeron--nucleon form factor was not implemented correctly. This results in a somewhat larger prediction cross section, but a factor of up to $\sim 2$, although as we will see the predicted rate remains negligible. 

\begin{table}
\begin{center}
\begin{tabular}{|c|c|c|c|}
\hline
& QCD, incoherent & QCD, mixed & QCD, coherent\\
 \hline
 ATLAS~\cite{ATLAS:2020hii} selection&0.8&0.2&0.07\\
  \hline
 ATLAS~\cite{ATLAS:2020hii} selection, $\alpha < 0.01$&0.2&0.05&0.06\\
  \hline
 CMS~\cite{CMS:2024bnt} selection&1.0&0.2&0.08\\
   \hline
 CMS~\cite{CMS:2024bnt} selection, $\alpha < 0.01$&0.2&0.05&0.06\\
\hline
\end{tabular}
\end{center}
\caption{Predicted percentage gluon--initiated (QCD) contributions the coherent `light--by--light' scattering final state, within the ATLAS~\cite{ATLAS:2020hii} and CMS~\cite{CMS:2024bnt}  event selections. Results with and without the acoplanarity cut $\alpha< 0.1$, are shown for comparison, although  for both measurements this cut is applied.}  \label{tab:lbyl_qcd}
\end{table}

In Table~\ref{tab:lbyl_qcd} we show the predicted QCD--initiated cross section for the LybL diphoton final state. We note that `incoherent' corresponds here to the case that both ions interact incoherently, which is in contrast to the nomenclature used above in the PI case, where this referred to the case that one ion interacted incoherently; here, we use the label `mixed' to refer to this case. We can see that the fractional contributions within the ATLAS and CMS event selection are at the sub--percent level, and indeed the permille level once the acoplanarity and diphoton transverse momentum cuts are imposed. The fully incoherent case is predicted to give the largest contribution, as we might expect given the short--range nature of the QCD interaction and consequent suppression from requiring the ion to interact as a fully coherent QCD source. The mixed and coherent are somewhat comparable after the full event selection is applied. In comparison to~\cite{Harland-Lang:2018iur} the predicted coherent and incoherent cases are indeed somewhat larger, as expected form the discussion above.

 \begin{figure}
\begin{center}
\includegraphics[scale=0.63]{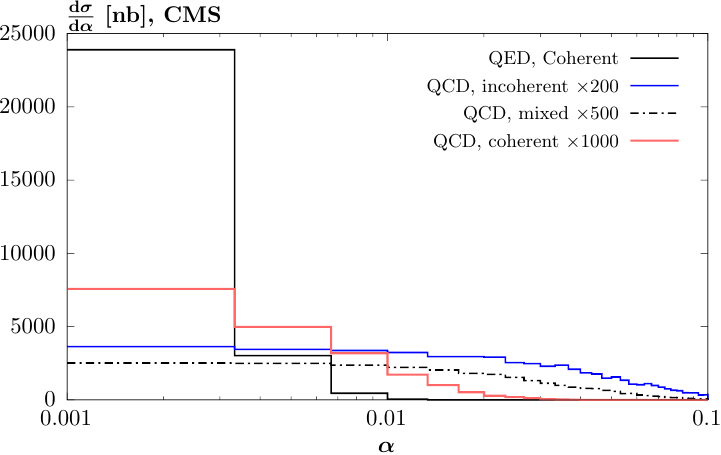}
\caption{Predicted coherent photon--initiated (QED) and gluon--initiated (QCD) `light--by--light' scattering cross sections, differential with respect to the diphoton acoplanarity, within the CMS~\cite{CMS:2024bnt} event selection. The QCD contributions are scaled as indicated in the legend for ease of comparison.}
\label{fig:CMSlbyl_aco_qcd}
\end{center}
\end{figure}

The predicted acoplanarity distributions are shown in Fig.~\ref{fig:CMSlbyl_aco_qcd}. As expected, and already observed in~\cite{Harland-Lang:2018iur}, the QCD--initiated contribution has a rather broader acoplanarity tail than the LbyL scattering signal. This is however less evident in the purely coherent case, due to the coherence requirement restricting the initial--state transverse momentum transfer to be relatively low. Therefore, it is only the incoherent and mixed contributions that can it appears potentially explain the large acoplanarity backgrounds observed by ATLAS and CMS. Moreover, we can see that it is only after scaling these contributions up by significant factors that this is possible; in other words, these can potentially describe the shape of this background in acoplanarity, but not the size of it.

However, some care is possibly needed here. In particular, there are various important sources of uncertainty in the modelling of this process. Indeed it is explicitly commented on in the CMS analysis~\cite{CMS:2024bnt} that the `large uncertainties' in the modelling of this process justify the overall normalization being left free. Now, while it is certainly true that there are important sources of uncertainty here, as discussed in e.g.~\cite{Harland-Lang:2014lxa}, there are limits on this provided by measurements in proton--proton (and proton--antiproton) collisions at the LHC and Tevatron. With this in mind, it is worth examining whether the theoretical uncertainties in the modelling of the QCD--initiated process can indeed account for the discrepancy between the data and theory, under the assumption that this process is indeed the dominant source of the observed background. The key fact here is that the observed background ratio is found to be $\sim 50\%$ of the LybL component in the signal region, while from Table~\ref{tab:lbyl_qcd} we can see that the dominant incoherent component is expected to be at the permille level, i.e. two orders of magnitude lower than this. 

As we might expect, it is indeed very hard to justify such a discrepancy on the grounds of a theoretical uncertainty of this order. A principle element of this is the existing CDF measurement of exclusive diphoton production~\cite{CDF:2011unh}. In the original analysis it is observed that the Durham model predictions are in rather good agreement with this. Taking the latest \texttt{SuperChic} MC implementation, and with all shared theoretical ingredients with the heavy ion case (PDF choice, scale etc) kept the same, we predict a central cross section of $2.4$ pb, which is in excellent agreement with the CDF measurement of $2.5 \pm 0.6$ pb (where we have roughly added the experimental uncertainties in quadrature for brevity). The Tevatron c.m.s. energy here is 1.96 TeV, which is only a factor of just over 2 less than the nucleon--nucleon c.m.s. energy relevant for the ATLAS and CMS measurements. Hence, while there will clearly be some extrapolation uncertainty between these two energies, and otherwise differences from the particular cuts etc, this gives us very good grounds for believing that the underlying nucleon--nucleon cross section is well modelled. 

 \begin{figure}
\begin{center}
\includegraphics[scale=0.63]{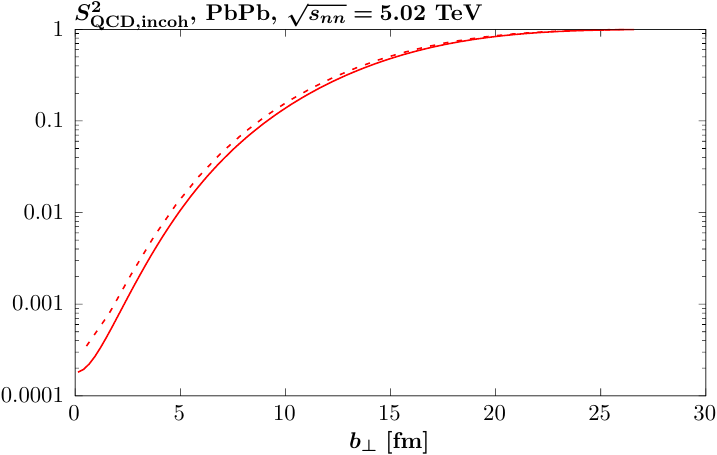}
\caption{Predicted survival factor for incoherent gluon--initiated diphoton production in PbPb collisions, at $\sqrt{s_{nn}}=5.02$ TeV, as a function of the nucleon--nucleon impact parameter, $b_\perp$. The dashed curve corresponds to the extreme and unrealistic case of dividing the nucleon--nucleon inelastic cross section by a factor of 2, and is shown for the sake of illustration.}
\label{fig:s2_incoh}
\end{center}
\end{figure}

Now, we  are interested here in the case of heavy ion collisions, and so there is certainly an additional source of uncertainty due to the modelling of the ion--ion cross section in terms of the nucleon--nucleon cross section, in essence due to the modelling of the underlying ion--ion survival factor.  Focussing on the case of purely incoherent production (although for the coherent and mixed cases the situation is similar), this survival factor suppression, which is given approximately by \eqref{eq:s2incoh}, is indeed rather significant, at the $\sim 10^{-4}$ level. This is demonstrated in an alternative manner in Fig.~\ref{fig:s2_incoh}, where the effective survival factor as a function of the nucleon--nucleon impact parameter is shown; the average suppression, which lies close to the $b_\perp \sim 0$ region, due to the short--range nature of the QCD interaction, can indeed be seen to be at this level. This plot then demonstrates the extent to which a significantly higher survival factor, and certainly one that is increased by the $\sim 2$ orders of magnitude required to match the ATLAS and CMS data, is  ruled out on rather general grounds. In particular, this is only achieved by going to impact parameters at the $\sim 5$ fm level or higher, which is clearly well outside the range of QCD. Another way to demonstrate this is shown by the dashed curve, which comes from the extreme and unrealistic case of dividing the nucleon--nucleon inelastic cross section by a factor of 2. In even this case, the survival factor in the low $b_\perp$ region increases somewhat, but clearly not sufficiently; this serves as a clear indication that any reasonably variation of the underlying model parameters is not going to provide the necessary source of uncertainty.

As a final comment on this, we note that in the original analysis of~\cite{Harland-Lang:2014lxa} the result of simply scaling the nucleon--nucleon cross section by a factor of $\sim A^2 R^4$, where the $R^4$ factor accounts for nuclear shadowing effects. If we simply scale the predicted nucleon--nucleon cross section by a factor of $A^2$, then for the CMS event selection we find that the predicted incoherent cross section is comparable in size to the coherent PI signal. Therefore, in order to match the observed CMS background would require essentially no suppression from ion--ion interactions. In terms of nuclear shadowing effects, we note that these are not included in Table~\ref{tab:lbyl_qcd}, in other words a free proton PDF is used throughout, but if an appropriate nuclear PDF is used than the predicted cross section are further reduced.

\section{Summary and Outlook}\label{sec:conc}

In this paper we have  for the first time provided a complete treatment of incoherent photon--initiated (PI) production in ultraperipheral heavy ion collisions, focussing on the dilepton and diphoton final states. In the former case we have compared to the ATLAS measurement~\cite{ATLAS:2022srr} of dielectron production and found that our predictions match the data very well. In the latter case we have shown that this does not provide the dominant contribution to the observed background to the purely coherent LbyL signal. We have in addition revisited the case of QCD--initiated diphoton production and have found that a similar conclusion holds, as we will summarise below.

The new modelling of incoherent PI production, and the updated modelling of the QCD--initiated contributions to the LbyL final state will be included in the \texttt{SuperChic} MC generator, and is publicly available at
\begin{center}
    \href{https://github.com/LucianHL/SuperChic}{https://github.com/LucianHL/SuperChic}.
\end{center}
 
We present below a summary of the implications of the above results for our understanding of the ATLAS~\cite{ATLAS:2020hii} and CMS~\cite{CMS:2024bnt} measurements of LybL scattering, and in particular the larger acoplanarity background to this process. We  recall that this is extracted using a template fit to the diphoton acoplanarity, making use of the fact that the signal is highly peaked at low values of acoplanarity and therefore in principle well separated from the background, which extends to higher values. The size of the background in the $\alpha < 0.01$ signal region is found to be rather large in terms of the number of events, comprising $\sim 50\%$ of the signal (somewhat less for ATLAS, and somewhat more for CMS). 
 
 Without considering any theoretical input, as a first point of comparison we can consider the measurement of dielectron production by ATLAS~\cite{ATLAS:2022srr}. This in particular applies a  similar (though not identical) template fitting technique to extract the coherent UPC signal, but also provides measurements of the cross section fraction due to the higher acoplanarity tail that is not described by the coherent signal modelling. Across the entire acoplanarity region this is found to be $\sim 4\%$ in the full fiducial region, and hence in the lower $\alpha < 0.01$ region would be lower still. 
 
We note  that the kinematic regions and c.m.s. energies are  similar between the dielectron and LybL measurements. Therefore,  based on the above information alone, it is clear that the source of high acoplanarity background in the LybL case must be dominantly of a different nature to that in the dielectron case. This has been explicitly verified  in this paper, where we have shown that the size and shape in acoplanarity of the dielectron background is well described by incoherent PI production, where one of the ions interacts incoherently and breaks up. The predictions for this background in the LybL case are as expected at the same order, and in fact after accounting for the precise kinematic cuts lower still. These therefore lie over an order of magnitude  below  the  background event fractions measured by ATLAS and CMS.

The question is then what differences there are between the dielectron and LybL processes that can lead to the differing size of the higher acoplanarity background in the two cases. One factor is that the final--state in the LbyL case can be couple (indirectly) to gluons, and hence be produced via a QCD--initiated mechanism at the same loop order as the PI case, while this is not the case for dielectron production. A second factor is that the diphoton and dielectron final--states themselves are different, such that different production topologies   enter in the incoherent PI case; for diphoton production it is of particular note that the LbyL signal proceeds via a loop--induced process and is therefore relatively suppressed. We have considered both possibilities here.

In particular, we have revisited the modelling of QCD--initiated production and find that no realistic modelling of this process, and account of the theoretical uncertainties in it can lead to a background of the size observed by ATLAS and CMS, which lie about 2 orders of magnitude above theoretical expectations. The size of the nucleon--nucleon cross section is in particular constrained by earlier CDF measurement of this process~\cite{CDF:2011unh} to be of the order of that in our theoretical predictions, such that to match the ATLAS and CMS observations would essentially require no additional suppression from the fact that most of these nucleon--nucleon interactions will lead to  ion--ion overlap and an inelastic final state. This is certainly at odds with the short range nature of the QCD interaction.

We have also considered for the first time the impact of the `quark emission' topology to the incoherent PI process, where the interacting quark in the incoherent ion emits the final--state photons directly. Given this bypasses the need for an intermediate loop as in the default LybL process, it enters at a lower order in $\alpha$ and hence might play a role; it is for this reason that the relative contribution could be enhanced relative to the dielectron case. However, on general grounds we find this is not expected to be the case, and we have verified this with an explicit calculation. Related to this, we have also  commented on the possibility of purely exclusive two--photon photoproduction off the nucleon. In general  this is expected to again be suppressed, being proportional to the valence generalized quark PDFs, although a remaining possibility remains that due to the related inelastic channel, a full analysis of which is left to future work.

Therefore, the reason for the increased size of the broad acoplanarity background in the LbyL case remains unclear. Given the size of this background, this is certainly an unsustainable situation with respect to future measurements. Even if the template fitting technique is viewed as a purely data--driven method, the accuracy of the extrapolation into the signal region and the dependence on this on the other kinematics of the process will clearly depend on this background being accurately modelled. As the current available models for the various components do not match the observed normalization of this background, it is hard to argue convincingly that this is the case. 

In the future, there are  approaches that can be used to mitigate this. An obvious possibility is study to LbyL scattering in the $0n0n$ ZDC class, for which the above incoherent (QCD and PI) mechanisms should not be present. If a reasonable high acoplanarity background persists for this ZDC class then it would certainly point to some other source of background being present. Further to this, an additional LHC measurement of diphoton production in $pp$ collisions would provide a cross--check of the so--far only measurement of this process, by the CDF collaboration, and hence of the modelling of the QCD--initiated background. In addition, providing measurements of the higher acoplanarity background, differential in the final--state kinematics, would assist in the modelling of this background. With such further analysis, we may hope to shed some light on this challenging issue.

\section*{Acknowledgements}

I thank Iwona Grabowska--Bold for initial interesting discussions and for motivating this study, and Valery Khoze and Misha Ryskin for reading through the manuscript and for their many useful comments. I thank the Science and Technology Facilities Council (STFC) part of U.K. Research and Innovation for support via the grant award ST/T000856/1.

 \appendix
 
 \section{Structure function inputs}\label{sec:sfinput}

In this appendix we describe the inputs for the inelastic and elastic structure functions entering the photon density matrix~\eqref{eq:rho} in Section~\ref{sec:nuccs}. For an ion beam we are only interested in the case of elastic photon emission, in which case the structure function $F_1$ is given in terms of the magnetic form factor of the ion, which is suppressed by $\sim Z^2$ with respect to the electric form factor contribution, $F_2$, and can be safely dropped. We in particular recall that for incoherent production we are considering the case where one ion continues to interact elastically and hence from \eqref{eq:aa_approx} this is suppressed by a factor of $\sim Z$ at the cross section level. That is, the magnetic form factor contribution remains parametrically suppressed with respect to incoherent production, as well as coherent. We therefore have that
\be
F_2^{\rm el, A}(x_{B,i},Q_i^2) = F_p^2(Q_i^2)G_E^2(Q_i^2) \delta(1-x_{B,i})\;, \qquad F_1^{\rm el, A}(x_{B,i},Q_i^2) = 0 \;.
\ee
Here  $G_E$ is the `Sachs' form factor of the proton. That is, we have included a term due to the form factor of the protons within the ion; numerically this has a negligible
impact, as the ion form factor falls much more steeply, however we include this for completeness. The ion form factor $F_p(Q^2)$ is given in terms of the proton density in the ion, and is calculated as described in~\cite{Harland-Lang:2018iur}.

For elastic emission from a proton  we have
\begin{equation}\nonumber
F_2^{\rm el, p}(x_{B,i},Q_i^2)=F_E(Q_i^2)\,\delta(1-x_{B,i})\;,\qquad F_1^{\rm el, A}(x_{B,i},Q_i^2) = G_M^2(Q^2_i)\;,
\end{equation}
where
\be
F_E(Q_i^2)=\frac{4m_p^2 G_E^2(Q_i^2)+Q^2_i G_M^2(Q_i^2)}{4m_p^2+Q^2_i}\;,
\ee
and $G_M$ is the magnetic `Sachs' form factor of the proton. For these, we use the fit from the A1 collaboration~\cite{A1:2013fsc}. Although elastic emission from a neutron is non--zero and can be accounted for following e.g. the approach of~\cite{Harland-Lang:2019pla} we have verified that this provides a negligible contribution and hence it can be safely omitted in the current study.

For inelastic emission, we use: CLAS data~\cite{CLAS:2003iiq} on inelastic  structure functions in the resonance $W^2 < 3.5$ ${\rm GeV}^2$ region, primarily concentrated at lower $Q^2$ due to the $W^2$ kinematic requirement; the HERMES fit~\cite{Airapetian:2011nu} to the inelastic low $Q^2 < 1$ ${\rm GeV}^2$ structure functions in the continuum $W^2 > 3.5$ ${\rm GeV}^2$ region; inelastic high $Q^2 > 1$ ${\rm GeV}^2$ structure functions for which the pQCD prediction in combination with PDFs determined from a global fit  provide the strongest constraint (we take the ZM--VFNS at NNLO in QCD  predictions for the structure functions as implemented in~\texttt{APFEL}~\cite{Bertone:2013vaa}, with the \texttt{MSHT20qed\_nnlo} PDFs~\cite{Cridge:2021pxm} throughout). All of the above inputs are precisely as in earlier studies, see e.g.~\cite{Harland-Lang:2020veo,Bailey:2022wqy}.

The latter discussion applies to the case of inelastic emission from a proton. For the neutron case we follow the approach of~\cite{Harland-Lang:2019pla}, namely for the resonant region we take the explicit CLAS fit to the inelastic neutron structure functions, while for the continuum region we reweight the structure functions by the charge weighted singlet of the neutron to the proton PDFs.

\bibliography{references}{}
\bibliographystyle{h-physrev}

\end{document}